# Constitutive Laws and Failure Models for Compact Bones Subjected to Dynamic Loading

M Pithioux, P Chabrand, M Jean
Laboratoire de Mécanique et d'Acoustique, CNRS, équipe MMCB, 31 ch. J. Aiguier
13402 Marseille, France

ABSTRACT:
Many biological tissues, such as bones and ligaments, are fibrous. The geometrical structure of these tissues shows that they exhibit a similar hierarchy in their ultra-structure and macro-structure. The aim of this work is to develop a model to study the failure of fibrous structures subjected to dynamic loading. The important feature of this model is that it describes failure in terms of the loss of cohesion between fibres. We have developed a model based on the lamellar structure of compact bone with fibres oriented at 0°, 45° and 90° to the longitudinal axis of the bone, and have studied the influence of the model parameters on the failure process. Bone porosity and joint stress force at failure were found to be the most significant parameters. Using least square resolution, we deduced a phenomenological model of the lamellar structure. Finally, experimental results were found to be comparable with our numerical model.

## INTRODUCTION

Biological materials such as bones and ligaments are fibrous structures. When considering trauma which occur accidentally or during sports, the fibrous structures of cortical bones are exposed to dynamic tearing, damage and failure mechanisms which are induced by shocks. Quite a lot of research has been carried out on the mechanical behaviour of hard and soft tissues when exposed to quasi-static forces, but the damage and failure caused by dynamic loading has not drawn the attention of many authors. A few have carried out experiments using micro-macro techniques. We can refer to the homogenisation method developed by (5), and the thermodynamic approach proposed by (3). Other authors have focused their work on the study of the mechanical behaviour of bone microstructure (1, 2, 6, 8, 14, 20, 21, 24, 25). These models cannot however be used to study bone failure. The aim of this work is to develop a specific model to study the behaviour of fibrous structures subject to dynamic forces until failure. We assume that the failure of such materials is caused by the loss of cohesion between fibres. These fibres are considered to be the elementary structural components with the lowest failure characteristic. Failure does not generally occur in such materials as a result of the fibres breaking, but rather because they lose their structural stability in an avalanche process of cohesive failure. We established this model by considering longitudinal as well as transversal joints between fibres. We then used a multi-scale method to deal with the various hierarchical levels of the structure. The purpose of this study is to show that numerical experiments, using a model of the cohesive composite material comprising elastic fibres, may help to understand the brittle behaviour of these tissues when they are subjected to dynamic tensile loading. Our approach is based on the finite element model described below, where the behaviour of the joints between the elastic fibres constituting the model are described by unilateral contact, friction and cohesive laws (12, 13). Since many of parameters in this model (e.g. structure heterogeneity, porosity, geometry, etc.) can influence the failure behaviour, we analysed a number of parameters to identify which ones have the greatest influence on failure of compact bones. This led us to propose laws concerning the structural constitution including these parameters. Finally, we compared our numerical results with experimental tests.

## I. STRUCTURAL ANALYSIS OF BONES

A general overview of the specific nature of the hierarchical structure of bone is helpful in understanding why we wanted to develop this fibrous model.
The constitution and geometric structure of bones are highly complex (7, 26). This means that they can be considered as a composite material. From a geometrical point of view at the lamellar level, the structural pattern of bones shows strong similarities with tropocollagen (24, 25) (Figure 1). The main deformation mechanisms can occur at five different levels:

- In the tropocollagen molecule and the microfibril (length = $2800\,\overset{0}{A}$), elongation mechanisms set the molecule profile (2, 18, 24, 25). The authors (4, 17) have proposed mechanical models based on the waviness of the microfibrils. At this level, the microfibrils were found to be elastic (10, 19, 20, 23).

- In the fibrils (length = 10 μm to 100 μm), the arrangement of overlapping and gap regions has been described on the basis of the Hodge Petruska model (11, 24, 25). The joints in the bone structure consist of a mineral hydroxyapatite component shown to exhibit elastoplastic behaviour (9, 14, 15, 24, 25).

- *The fibre* is a geometrical assembly of disjointed fibrils embedded in an amorphous substance. When fibres are loaded, the pressure applied causes fluid to be exuded from the tissue, as the result of which the material exhibits viscous behaviour (16).

- The *fibre-lamella* assembly is made up of fibres which are all oriented in the same direction. There are several different kinds of lamellae, oriented longitudinally, transversally and obliquely, *i.e.* forming an angle of 0°, 45° and 90° with the longitudinal axis of the structure.

- The *lamella-osteonal and osteonal-bone* assembly cannot be precisely described in terms of fibre interactions.

## II FIBROUS MODEL AND EXPERIMENTAL METHOD

Our model is based on the biological fibrous structure, and the failure process is assumed to be caused by successive cohesive failure events occurring at different scales and aggregating so as to generate micro-voids or micro-cracks, that well up into aggregates generating voids or cracks on a larger scale. These processes could be studied statistically. There remains the possibility of exploring an assembly of model-fibres linked together with appropriately distributed cohesive forces. Given the fibrous structure of the model, the ability to bear tensile stress results from two classes of cohesive forces. These are: (i) the fore end of each fibre on the aft end of the previous fibre (head to tail or longitudinal cohesive force); and (ii) the flanks of neighbouring fibres (flank to flank or transversal cohesive force). The principal component of the head to tail forces is tensile, while the principal component of the flank to flank forces is a shear force. In this study, tensile tests in the direction of the fibres were taken into account. The constitutive laws were therefore developed from a fibrous model subjected to tensile traction applied in the direction of the fibre axes (this model could be a lamellar structure). Because of the dynamic kinematics involved, damage was assumed to be preponderant, and viscosity was neglected. The following fibrous model (cf. Figure 2.a) consists of an assembly of fibres, in which joints are formed longitudinally and transversally between the fibres (cf. Figure 2.a and 2.b).

## II.1. The Joint Model

In this model, interactions between fibres are described using unilateral contact conditions and Coulomb's friction coefficient. Let n be the normal unit vector to the contact interface. The normal relative displacement between two contacting fibres is then defined by $q_N = q.n$. decomposing the contact force vector R into the normal ($R_N$) and tangential ($R_T$) components gives: $R = R_N.n + R_T$.

In the cohesive model, the unilateral constraints (Signorini conditions) are written as follows (1):

$$q_N \geq 0,\ R_N + l \geq 0,\ (q_N, R_N + l) = 0 \qquad (1)$$

where l is the longitudinal cohesive threshold.
The joint model characterises a cohesive Mohr Coulomb law (cf. Figure 3.a) which corresponds to a translation of the Coulomb cone.
The corresponding law for friction with adhesion in terms of the threshold μl is given by relations 2.a, 2.b and 2.c:

$$U_T = 0 \quad \Rightarrow R_T \in\ ]-\mu\,(R_N + l)\ ;\ \mu\,(R_N + l)[ \qquad (2.a)$$

$$U_T < 0 \quad \Rightarrow R_T = -\mu\,(R_N + l) \qquad (2.b)$$

$$U_T > 0 \quad \Rightarrow R_T = \mu\,(R_N + l) \qquad (2.c)$$

Where μl is the transversal cohesive breakdown threshold.
This cohesive law can result in two possible situations occurring at the contact interfaces:

(i) If ($R_N$, $R_T$) lies inside the Coulomb cone (point A in Figure 3.a for example) the fibres stick together and there is no relative displacement
(ii) If ($R_N$, $R_T$) lies on the cone boundary (point C for example in Figure 3.a), the cohesive status is lost and friction occurs between the fibres, which corresponds to local damage (i. e. the joints between the fibre components are broken)

The static sliding coefficient μ is defined by $\mu = tg\alpha = L/l = f_2/f_1$ (cf. Figure 3.a).
The cohesive thresholds differ between longitudinal ($f_1$) and transversal ($f_2$) joints and are related to maximal head to tail and flank to flank cohesive forces (Figure 3.b).

When the joints are broken, the classical unilateral Signorini conditions (3) involving Coulomb's friction (4.a, 4.b and 4.c) give:

$$q_N \geq 0\ ;\ R_N \geq 0;\ (q_N, R_N) = 0 \qquad (3)$$

$U_T=0 \quad \Rightarrow R_T \in ]-\mu_f R_N ; \mu_f R_N[$ (4.a)
$U_T<0 \quad \Rightarrow R_T = -\mu_f R_N$ (4.b)
$U_T>0 \quad \Rightarrow R_T = \mu_f R_N$ (4.c)
where $U_T$ is the sliding celerity and $\mu_f$ is the Coulomb's friction coefficient.

In order to account for the heterogeneity of the material, we applied a stochastic process to choose the cohesive thresholds for the model. This introduced local defects into the structure.

**II.2 Dynamic Equation and Boundary Conditions**

Assuming the fibres exhibit elastic behaviour, the principle of virtual power was used to obtain the dynamic equation, the discrete form of which leads to the following system:

$$M\ddot{q} + Kq = F + r \qquad (5)$$

where M is the mass matrix, K the rigidity matrix, $\ddot{q}$ (resp q) the nodal acceleration vector (resp displacement), F is the external force vector and r the contact force vector.

The boundary conditions must account for the behaviour of the surrounding material. On the lateral sides, the displacements were taken to be zero for all left-hand side nodes (Ux=0) (Figure 4). Though several different boundary conditions may be applied to the upper and lower sides, they were chosen to specify $\sigma_{yy} = 0$ and identical displacements in the y displacements of the upper and lower sides ($q_y^{sup} = q_y^{inf}$). These conditions are characteristic of periodic boundary conditions describing the effect of the material surrounding the sample in a tensile experiment. For all right-hand side nodes, only displacements along the x axis are specified:

$$q_x = \dot{\varepsilon}_{xx}(L \,.\, t) \qquad (6)$$

In (6) $\dot{\varepsilon}_{xx}$ is the specified deformation rate, L is the length of the fibre and t the time. The boundary conditions chosen limit swelling effects but the Poisson effect is free to act on the structure.
To reduce side effects, we have introduced a new set of variables in which the new variable u is defined by:

$$q(x,t) = u(x,t) + t\left[\dot{A}\right](x) \qquad (7)$$

$$\text{with } \dot{A} = \begin{bmatrix} \dot{\varepsilon}_{xx} & 0 \\ 0 & \dot{\varepsilon}_{yy} \end{bmatrix}$$

Where $\dot{\varepsilon}_{xx}$ and $\dot{\varepsilon}_{yy}$ are the constant specified deformation rate.
The discrete form of the dynamic equation can then be written as:

$$M\ddot{u} + Ku = -K\dot{A}t + F + r \qquad (8)$$

The boundary conditions become:
We specify $\sigma_{yy}$=0 with similar displacement of the upper and lower sides ($u_y^{sup} = u_y^{inf}$).

The advantage of such a variable change is that it gives a "smooth numerical" problem.
When a transformation of the discrete time domain is performed, an elementary subinterval $]t_I, t_{I+1}]$, of length h is considered. The principal consideration developed in this transformation is that discrete variables do not necessarily have to be defined at a specific time within this interval. In any case, the impact times are usually unknown, or it can be difficult or costly to approximate or isolate when simultaneous contacts occur.
Let $t_I$ be the time at the previous increment and $t_{I+1}$ the time of the current load in progress. $\dot{q}(I)$ denotes an approximation of $\dot{q}(t_I)$ and $\dot{q}(I+1)$ an approximation of $\dot{q}(t_{I+1})$. We use the same notation for the other parameters.
The discrete dynamic equation is solved using an implicit Euler method with the schema:

$$\begin{cases} h\ddot{q}(I+1) = \dot{q}(I+1) - \dot{q}(I) \\ q(I+1) = q(t_I) + h(\dot{q}(I+1) - \dot{q}(I)) + h\dot{q}(I) \end{cases} \quad (9)$$

At time $t_{I+1}$, the equation (5) can be rewritten as follows:

$$\begin{cases} \dot{q}(I+1) - \dot{q}(I) = w[-hK(q(I) + h\dot{q}(I)) + hF(I+1) + hr(I+1)] \\ \text{with } w = \left(M + h^2K\right)^{-1} \end{cases} \quad (10)$$

The “Non Smooth Contact Dynamic Method” presented in (13) is used to deal with the frictional contact problem.

## II.3 Finite Element Modelling

To analyse the development of failure in structural bone tissue, the elementary volume was chosen to represent the length of a lamella. This study can also be applied to a microfibril, a fibril or a fibre. A representative volume was obtained by assembling 21 fibres longitudinally in 30 successive layers. To study the effect of the geometrical pattern, we also considered a structure constituted from the same number of fibres but made with 90 successive layers of 7 longitudinal fibres each. We also studied the effect of the geometry by doubling the number of fibres, modelling 42 fibres longitudinally in 60 successive layers. With the cohesive model, each fibre was represented by a 2D elastic model-fibre (Figure 5). Each model-fibre was composed of eight T3-triangle linear finite elements (Figure 5). Cohesive frictional contact forces were exerted between model-fibres and, for numerical purposes, were assumed to be concentrated at specific midpoints (denoted in figure 5 by tiny circles). Three types of lamellar structure in compact bone were simulated, with fibres oriented longitudinally, obliquely and transversally; *i.e.* at 0°, 45° and 90° with respect to the longitudinal axis of the structure.
Several authors (1, 6, 8, 14, 19, 20, 23, 24) have measured and analysed the mechanical properties of collagen and hydroxyapatite (Young’s modulus, Poisson's ratio and density). In accordance with these works, the fibres were assumed to be elastic with a Young's modulus E between 11GPa and 16GPa. The Young's modulus considered was an average of the collagen and hydroxyapatite modulus because in our model-fibre the mineral and collagen components were considered as a single equivalent material. The density d of the fibres was $2\ 10^3$Kg/m$^3$ and the Poisson ratio $\nu$ was 0.33. The cohesive forces of the joints correspond to the plasticity values of the hydroxyapatite mineral.
$f_l$ is taken to be the cohesive head-tail stress for each layer and $f_2$ the cohesive shearing stress for each layer interface (Figure 6). In order to compare different responses obtained with a number of values of the ratio $f_l/f_2$ with a given global failure threshold, the resulting force $f_{res}$, was defined from:

$$f_{res} = (N-2)\ f_l + 2\ f_{\partial l} + (N-1)\ f_2 \quad (13)$$

with $f_{\partial l}$ the head tail force of the boundary fibres and N the number of layers

For the above relations, it was assumed that the boundary fibres are subjected to shear cohesive forces exerted by the neighbouring material equal to $f_2$. For the sake of simplicity and numerical processing, this force was included in the head tail force of the boundary fibres $f_{\partial 1}$ so that $f_{\partial 1} = f_1 + f_2$.
We obtain:
$f_{res} = N f_1 + (N+1) f_2$ (14)
Considering the height l and the length L of the fibrous structure (a unit value being assumed for thickness), the above formula can be written:

$$\frac{f_{res}}{Nl} = \frac{f_1}{l} + \frac{1}{2}\frac{N+1}{N}\frac{f_2}{L/2}\frac{L}{l} \qquad (15)$$

Equation (13) can be rewritten, introducing $\sum_{res}$ , the equivalent tensile stress, $\sigma_1 = \frac{f_1}{l}$, the head tail cohesive stress and $\sigma_2 = \frac{f_2}{L/2}$, the equivalent shear cohesive flank-flank stress:

$$\sum\nolimits_{res} = \sigma_1 + \frac{1}{2}\left(1 + \frac{1}{N}\right)\frac{L}{l}\sigma_2 \qquad (16)$$

Since $f_1$ and $f_2$ cannot exceed the maximum values of the cohesive threshold, the above formula (16) shows clearly that the stress-strain response curve depends on the $f_1/f_2$ ratio. If threshold values are measured as stresses, the ratio $\sigma_1/\sigma_2$ becomes a significant parameter together with the aspect ratio L/l.

**II.4 Experimental Tensile Method**

We assume that the most relevant parameters relating to the failure of compact bones need to be studied at the microstructure level. Using an X-ray scanner and a microscope, we analyse the sample density and porosity of the structure. We use an X-ray scanner ND8000 of the Laboratoire de Mécanique et d'Acoustique (Desrues J., et al., [1996], Bonnenfant D., et al. [1998]). X-ray scanner images were recorded on different sections of the bone samples. Sections were taken every 10 mm on a sample 110 mm long.
We then analyse precisely the density variations in the section where the density was found to vary slightly through the lamellar structure of all the bones studied (around 1.9 $10^3$ kg/m$^3$). In the osteonal structure, the density varies between 1.8 $10^3$ kg/m$^3$ and 2.1 $10^3$ kg/m$^3$. In parallel, a microscope was used to analyse the porosity of the bone structure.

Samples were taken from ten fresh bovine femoral bones. Two samples from each bone were cut in areas where the density varied slightly (around 1.9 $10^3$ kg/m$^3$) in the anterior lateral and anterior medial. These samples have a lamellar structure. Tensile experiments were developed using an INSTRON machine (figure 7) used at various velocities (0.5 mm/min to 50 mm/min) with 500 kg tensile force (21). Human femoral bones were used to make samples.

The epiphyses were cut so as to focus only on compact bone. Test-pieces were obtained by first cutting the bones in the axial direction and then removing the marrow from each specimen. Samples were then machined digitally, and shaped as shown in Figure 8.a, because bone is brittle when undergoing failure. Lastly, a gauge was positioned to measure the local deformation of the sample during the failure process (Figure 8.b). The local mechanical properties of compact bones were deduced where failure occurred. Stress-strain curves are obtained (22).

## III RESULTS: EFFECT OF MODEL PARAMETERS

Since we modelled dynamic traction experiments, the deformation mechanisms had a traction wave effect which was reflected on the left-hand side where zero displacement was specified. The plastic deformations observed in joints may have resulted from these effects.

### III.1 Effect of Structure Heterogeneity and Dimension

Calculations were developed for fibres oriented at 0°, 45° and 90° with respect to the longitudinal axis of the bone. First, in order to introduce heterogeneity into the structure, we defined an interval within which the longitudinal (traction thresholds) and transversal thresholds (shearing thresholds), chosen with a stochastic process, could vary. Second, the three geometries (21 longitudinal fibres in 30 successive layers, 7 longitudinal fibres in 90 successive layers and 42 longitudinal fibres in 60 successive layers) were tested. The force-strain curves showed that structure heterogeneity (Figure 9) and dimension (Table 1) had no effect on the value of the failure force. Oscillations showed propagation of tensile-compressive waves.
The force-strain curves (Figure 9) could be divided into two parts:
- The first part was an elastic linear region from which the values of the Young's modulus of the structure could be found.
- The second part was a damage region up to very brittle failure.

### III.2 Effect of the Scattering Parameter

In this case, computations were performed with fibres oriented at 0°, 45° and 90° with respect to the longitudinal axis of the bone. The length of the interval of the traction and shearing thresholds chosen under these conditions varied (it represents the scattering parameter). The scattering parameter influenced failure for all three fibres orientations: structure fractures occurred more rapidly with longer intervals (Figure 10).

### III.3 Effect of the Ratio $f_1/f_2$

Different assemblies of model-fibres, linked together with appropriately chosen cohesive forces, were then explored. The fibrous structure of the model was considered and the ability to bear tensile stress was found to be due to two classes of cohesive forces: forces exerted by the fore end of a fibre on the aft end of the proceeding fibre ($f_1$); forces exerted flank-to-flank between neighbouring fibres ($f_2$). In the context of tensile tests, $f_1$ forces were mainly tensile forces while $f_2$ forces were mainly shearing forces; the ratio $f_1/f_2$ was thus a significant mechanical parameter. We observed that the ratio $f_1/f_2$ had no effect on the failure force of the structure when fibres were oriented at 0° in comparison with the longitudinal axis of the bone (Figure 11). This parameter did however have an important effect on structure failure when the fibres were oriented at 45° and 90°, particularly when the longitudinal joints were stronger than the transversal joints (Figure 12).

## IV CONSTITUTIVE LAWS

From the previous results, we found that when the fibres are oriented at 0° with respect to the longitudinal axis, only the scattering parameter plays an important role in the failure process. We studied the evolution of the failure strain with respect to the scattering parameter. We used a least squares resolution method to find representative laws of compact bones at failure (Figure 13):

The phenomenological model found is:

$$\sigma = E\varepsilon \ \text{ if } \varepsilon \le \varepsilon_{ult} \tag{17}$$

$$\sigma = 0 \ \text{ if } \varepsilon \succ \varepsilon_{ult} \tag{18}$$

with $\varepsilon_{ult} = \varepsilon_0 \cdot d + \varepsilon_1$

d is the scattering parameter. $\varepsilon_0 = 1{,}8\ 10^{-6}$ and $\varepsilon_1 = 2{,}8\ 10^{-6}$

We used the same method for fibres oriented at 45° and 90° with respect to the longitudinal axis. In this case, failure depended on two parameters, the scattering and the ratio $f_1/f_2$. We deduced the constitutive laws of the structure:

$$\sigma = E\varepsilon \ \text{ if } \varepsilon \le \varepsilon_{ult} \tag{19}$$

$$\sigma = 0 \ \text{ if } \varepsilon \succ \varepsilon_{ult} \tag{20}$$

with $\varepsilon_{ult} = \varepsilon_2 \cdot d + \varepsilon_3 \cdot r + \varepsilon_4$;

d = scattering parameter; $r = f_1 / f_2$; $\varepsilon_2 = 1{,}8\ 10^{-6}$; $\varepsilon_3 = 7.8\ 10^{-6}$ and $\varepsilon_4 = -7.64\ 10^{-6}$

## V COMPARISON BETWEEN THE NUMERICAL MODEL AND EXPERIMENTAL RESULTS

The objective of this work was to show that a cohesive model analysing the failure of bones is in reality comparable to a model analysing macroscopic failure. In the model that we have developed, failure occurs suddenly, cohesive failure of a few joints triggering a step by step failure process in the neighbouring vicinity. Longitudinally and transversally connected joints were also damaged and the structure was broken in only a few steps (Figure 14). In the experimental results, failure also occurred very rapidly. One experimental failure pattern is illustrated in Figure 15. Qualitatively, the experimental and computational results were quite similar (Figure 14, 15). Quantitatively, a study taking into account scale effects and dispersion between one bone to another is in progress.

## VI. DISCUSSION AND CONCLUSION

To summarise, in this paper we propose a new method for investigating bone failure, following on from the development of earlier models of cortical bones. In our model, bones are described as fibrous structures in which deformation occurs in the fibrous elements and damage results from the failure of joints between the fibres. A finite element model is used to describe the behaviour of this fibrous structure, based on cohesive laws as applied to joint modelling. With the model, we have established how parameters such as scattering and the ratio of the traction and shear cohesion thresholds can strongly affect the description of the failure process. The cohesive model gives a physical description of the damage resulting from joint cohesive failure processes. The parameters describing the heterogeneity and the geometry do not affect the behaviour of the structure. The ratio between the longitudinal and the transversal failure stress joints (connecting the heads and flanks of adjacent fibres) and the scattering parameter, can be used to control the behaviour and the development of damage in the model. The physical meaning of the scattering parameter is given by the porosity of the material. These parameters are relevant for studying microstructure, and therefore macrostructure, failure. For the structure modelled here, statistics on the failure mode describing the development of the node status can be used to obtain an overall micro macro assembly. The most important feature of this model is that the failure behaviour of the structure depends on these parameters. With this approach, the internal value of a microscopic and therefore macroscopic phenomenological model dealing with the physics of the damage processes can be defined.

The qualitative results obtained with this finite element model are in complete agreement with those obtained experimentally for the lamellar structure of bone with an Instron device. This constitutes a first step towards describing a structural failure model in which damage occurs as the result of cohesive failure between the longitudinal and transversal joints.

A microscopic/macroscopic assembly is needed for studying bones. The method can be used to study the lamellar/osteonal assembly considering the lamella as the basic component of the osteonal. This method is a multi-scale method. Finally, this finite element model does not take into consideration the viscosity of the structure so as to emphasise the cohesive failure processes. We propose carrying out further calculations introducing for example a viscous component.

List of captions

Figure. 1: Assembly of the different hierarchical levels in the bone structure.
Figure. 2.a: Fibrous assembly
Figure. 2.b: Example of unit fibrous assembly projected onto a two dimensional plane.
Figure. 3.a: Mohr Coulomb cohesive law
Figure. 3.b: $f_1$ is the cohesive coefficient of the shear between two fibre layers. $f_2$ is the tensile cohesive coefficient for the fibres cohesion on the same layer.
Figure 4: Boundary condition
Figure 5: Elementary block 8 T3 and candidates for contact
Figure 6: Distribution of cohesive forces
Figure 7: INSTRON device
Figure 8: (a) isolated sample (b) sample with gauge
Figure 9: Strain-force curve with randomised thresholds
Figure 10: Effects of scattering on the force deformation curve. The scattering parameter was more important for 01 than for 02 and for 02 than for 03....
Figure 11: Effect of the parameter $f_1/f_2$ when fibres are at 0° in comparison with the longitudinal axis
Figure 12: Effect of the parameter $f_1/f_2$ when fibres are at 45° and 90° with respect to the longitudinal axis. When the parameter $f_1/f_2$ is equal to three, the longitudinal joints are stronger than the transversal joints, and the failure force is very different from the other cases.
Figure 13: Scattering-failure strain curve obtained with the model and when we apply a least squares solution
Figure 14: Numerical results obtained when the fibres are assumed to be oriented at 0° with respect to the longitudinal axis
Figure 15: Failure profile obtained with the INSTRON device. Sample has a lamellar structure with fibres all oriented at 0° with respect to the longitudinal axis
Table I: Failure stress for the three geometries

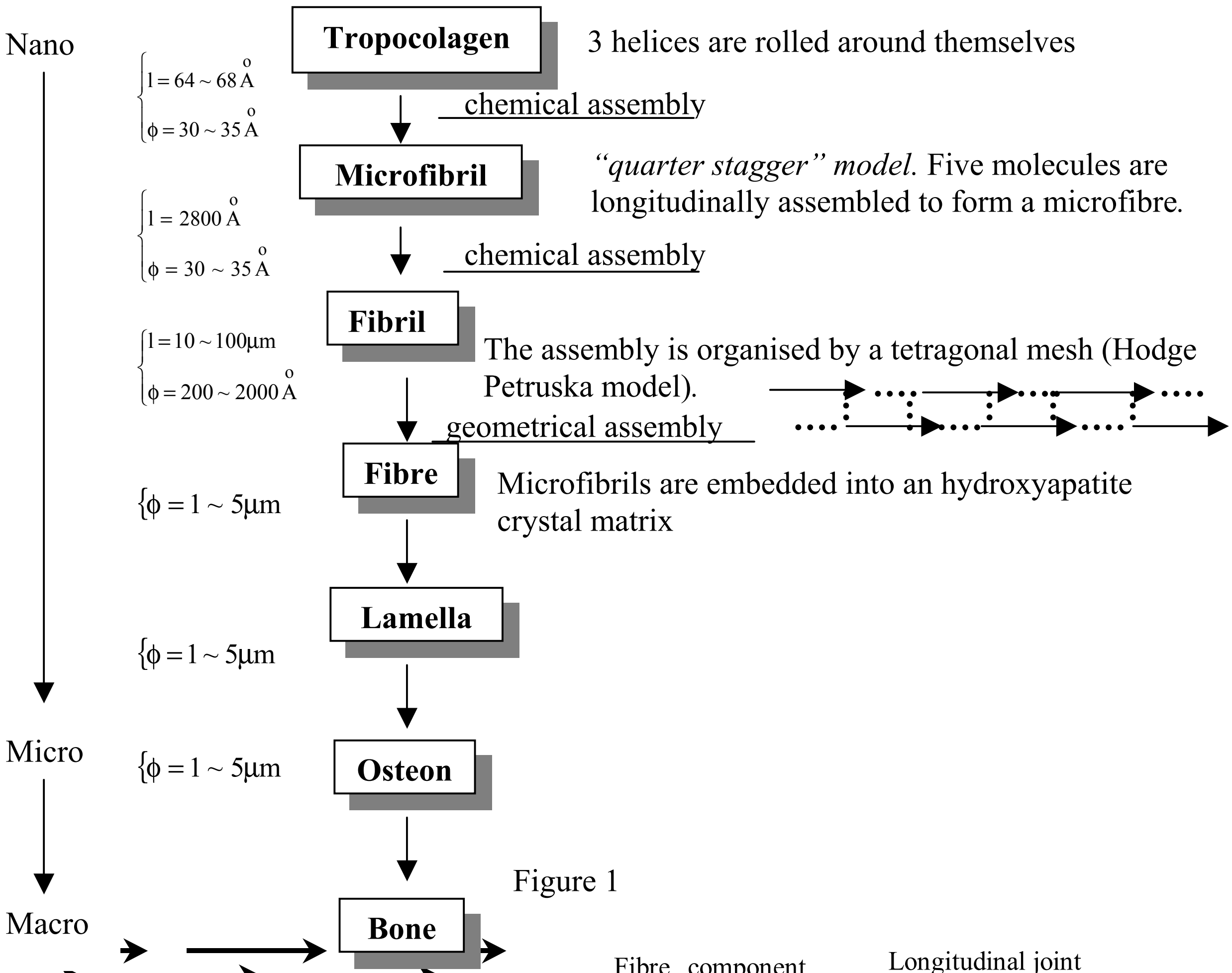

Figure 1

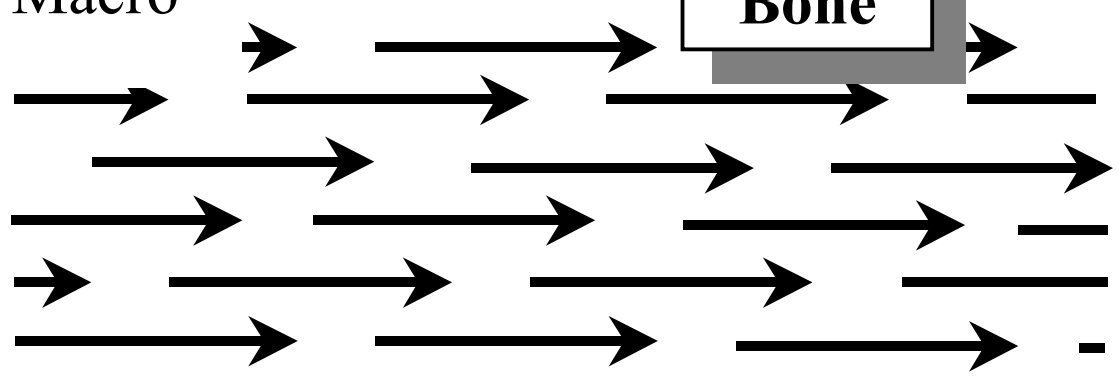

Figure 2.a

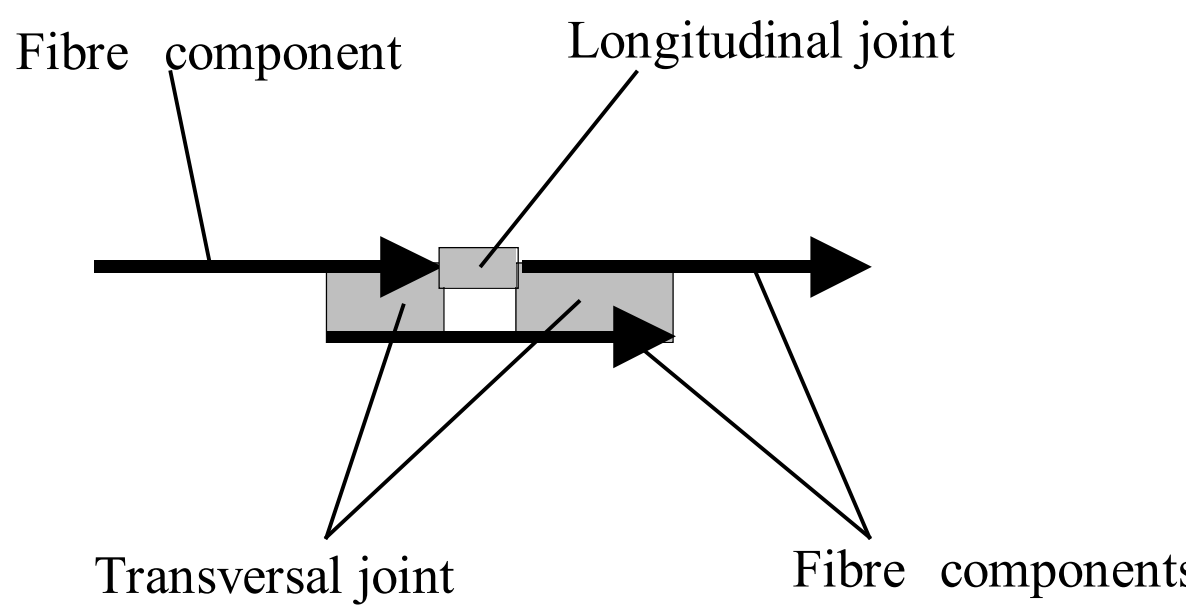

Figure 2.b

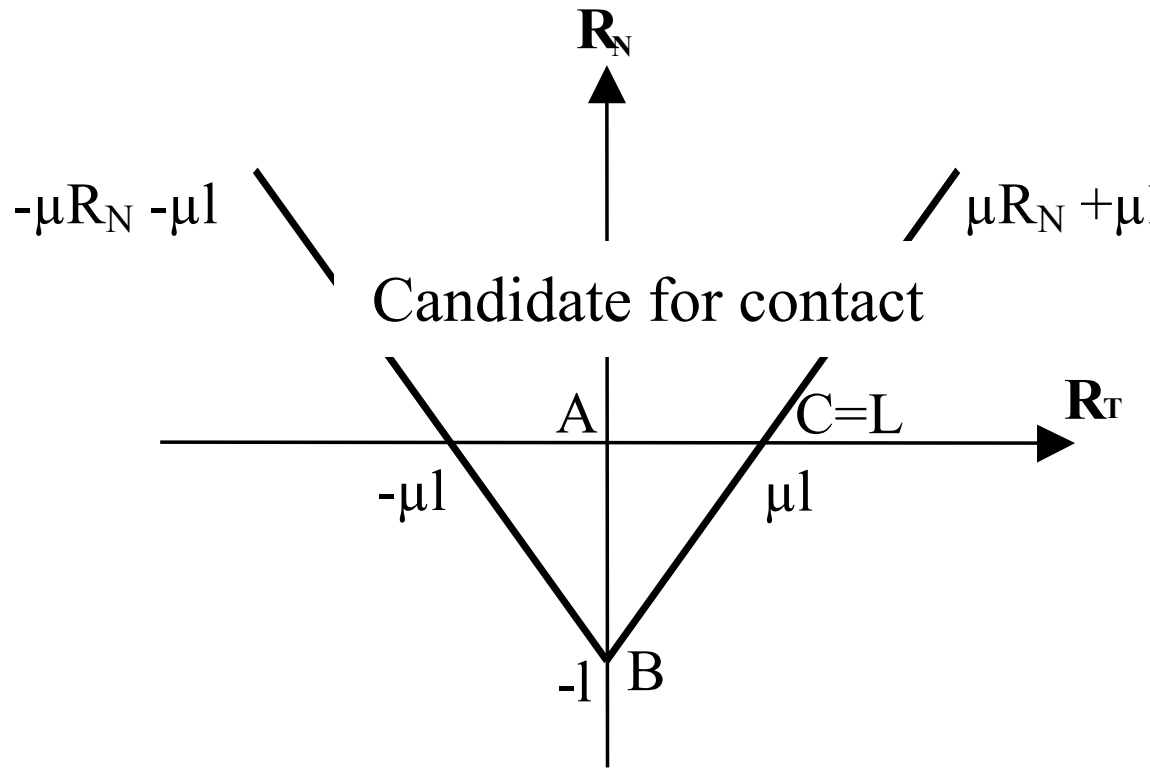

Figure 3.a

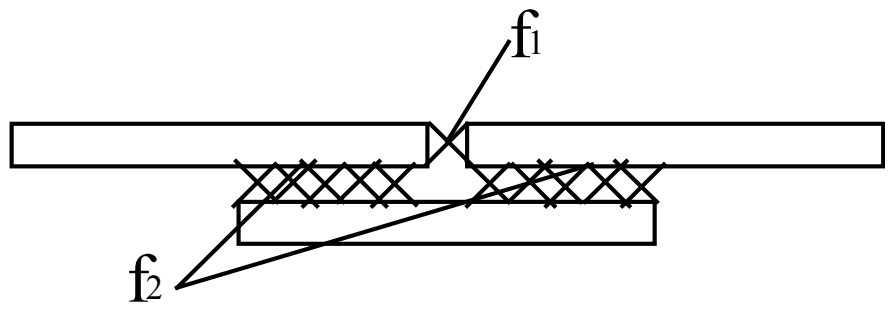

Figure 3.b

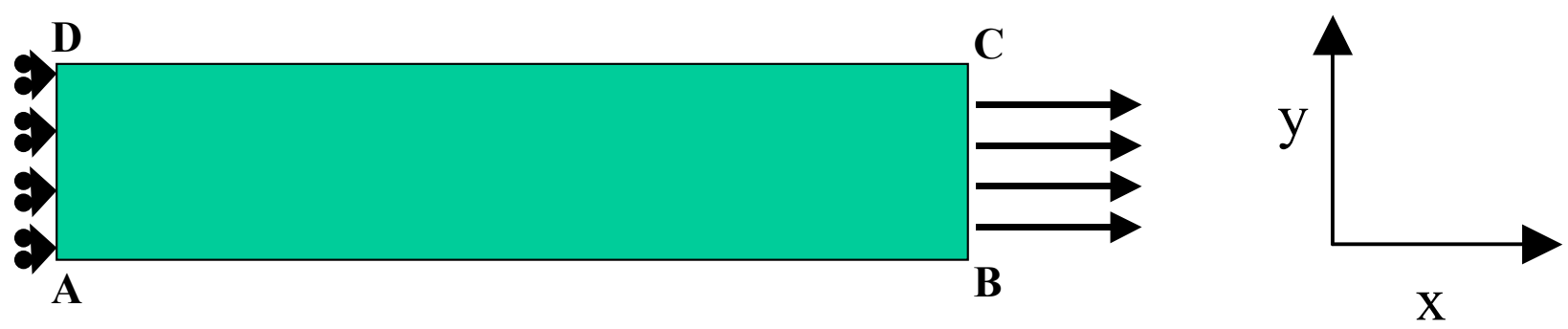

Figure 4

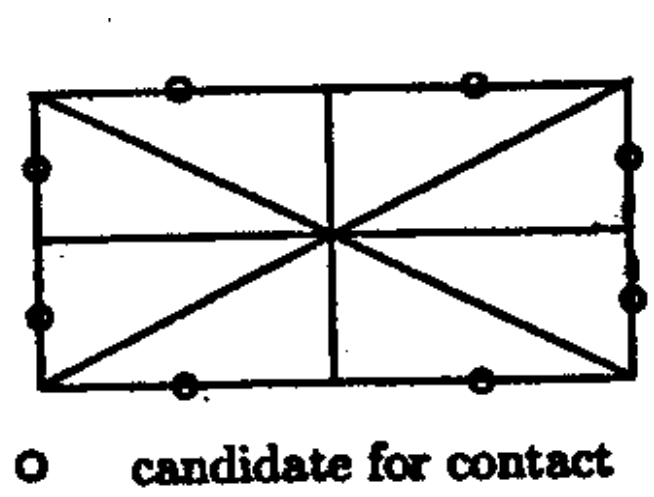

Figure 5

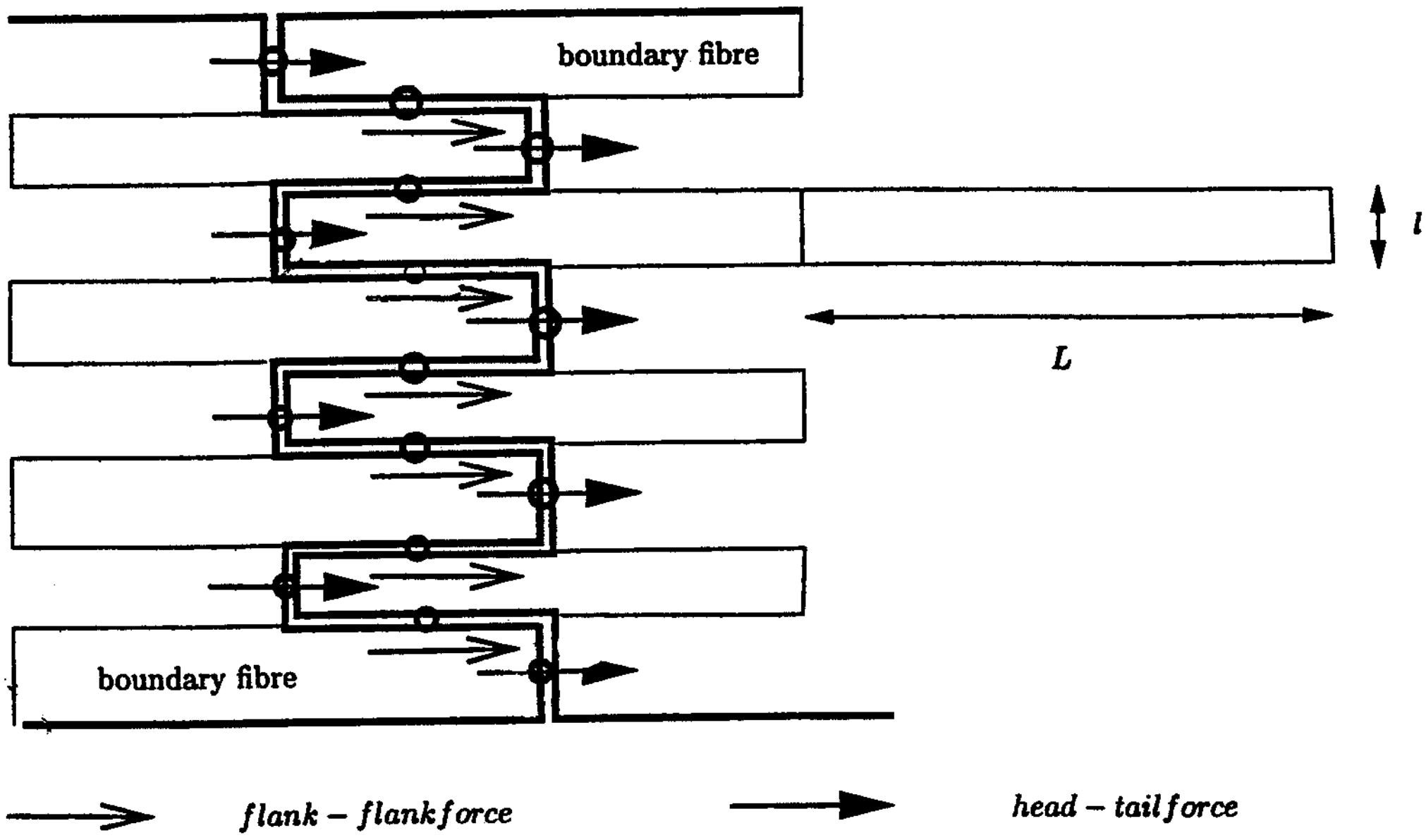

Figure 6

$f_2$, $f_1$,

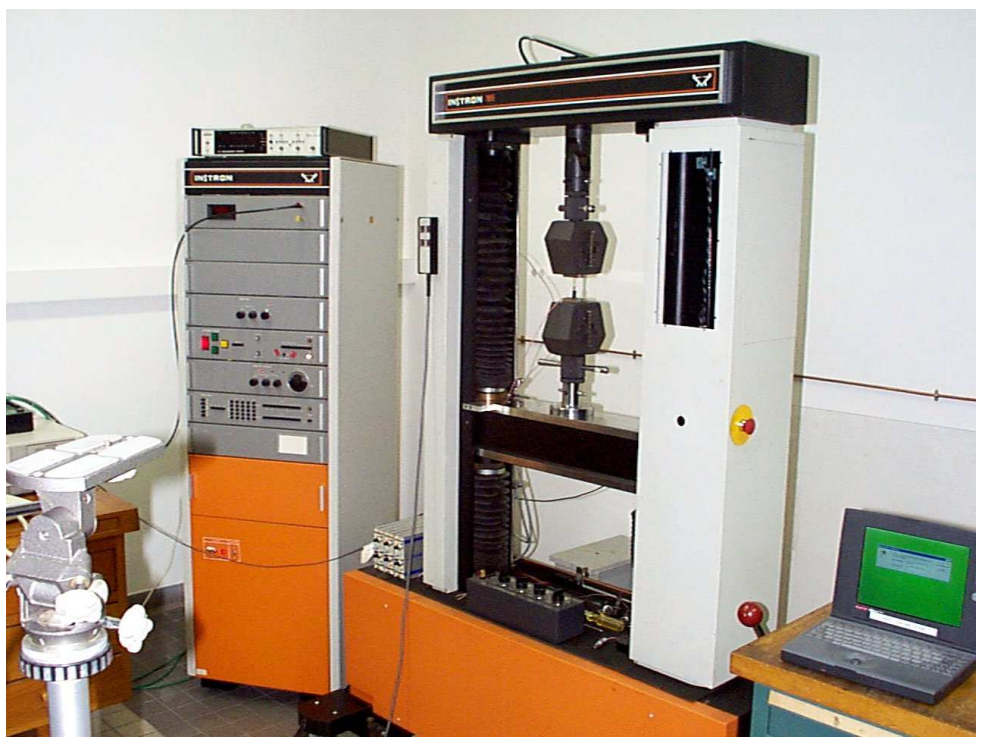

Figure 7

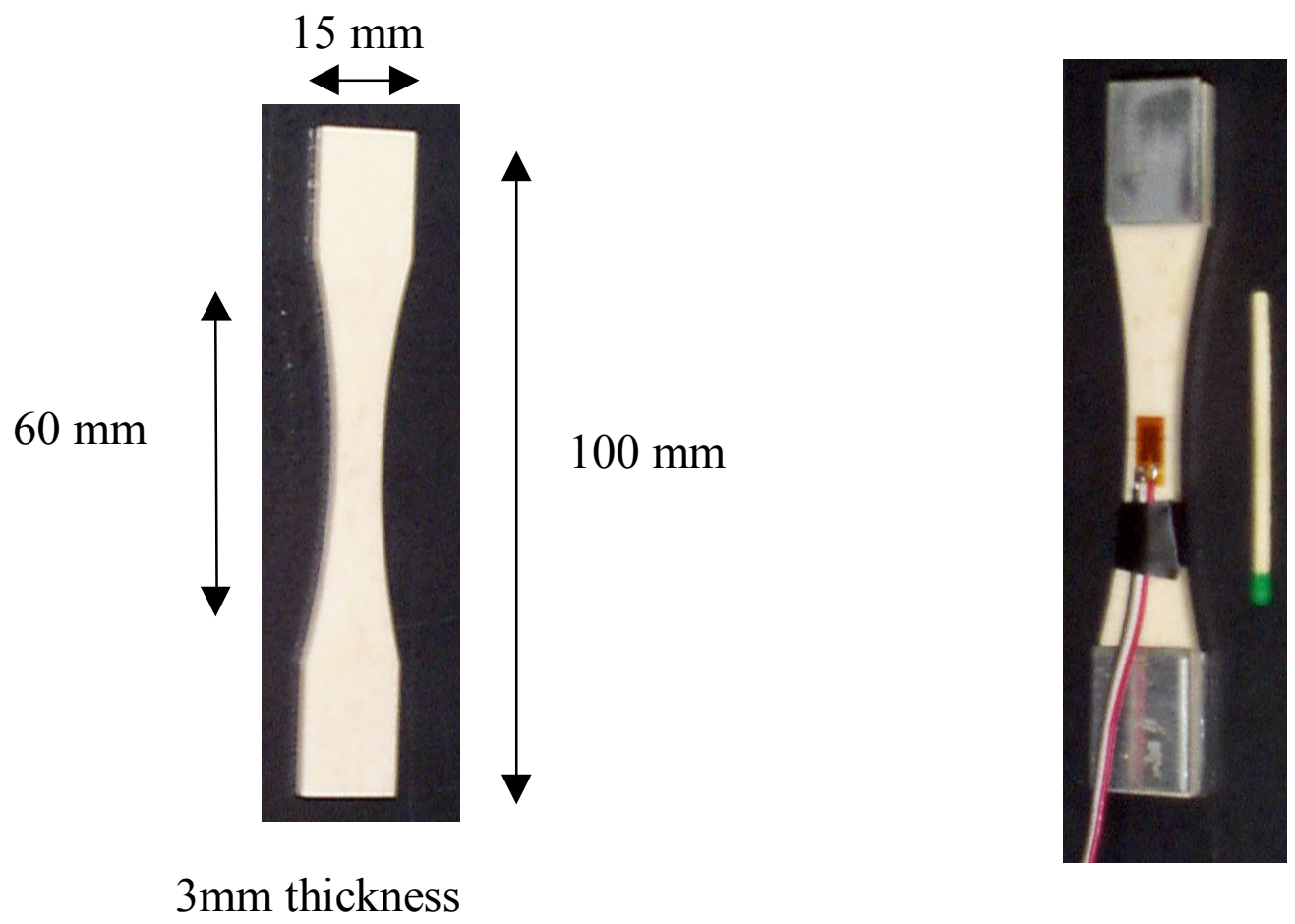

(a) (b)

Figure 8

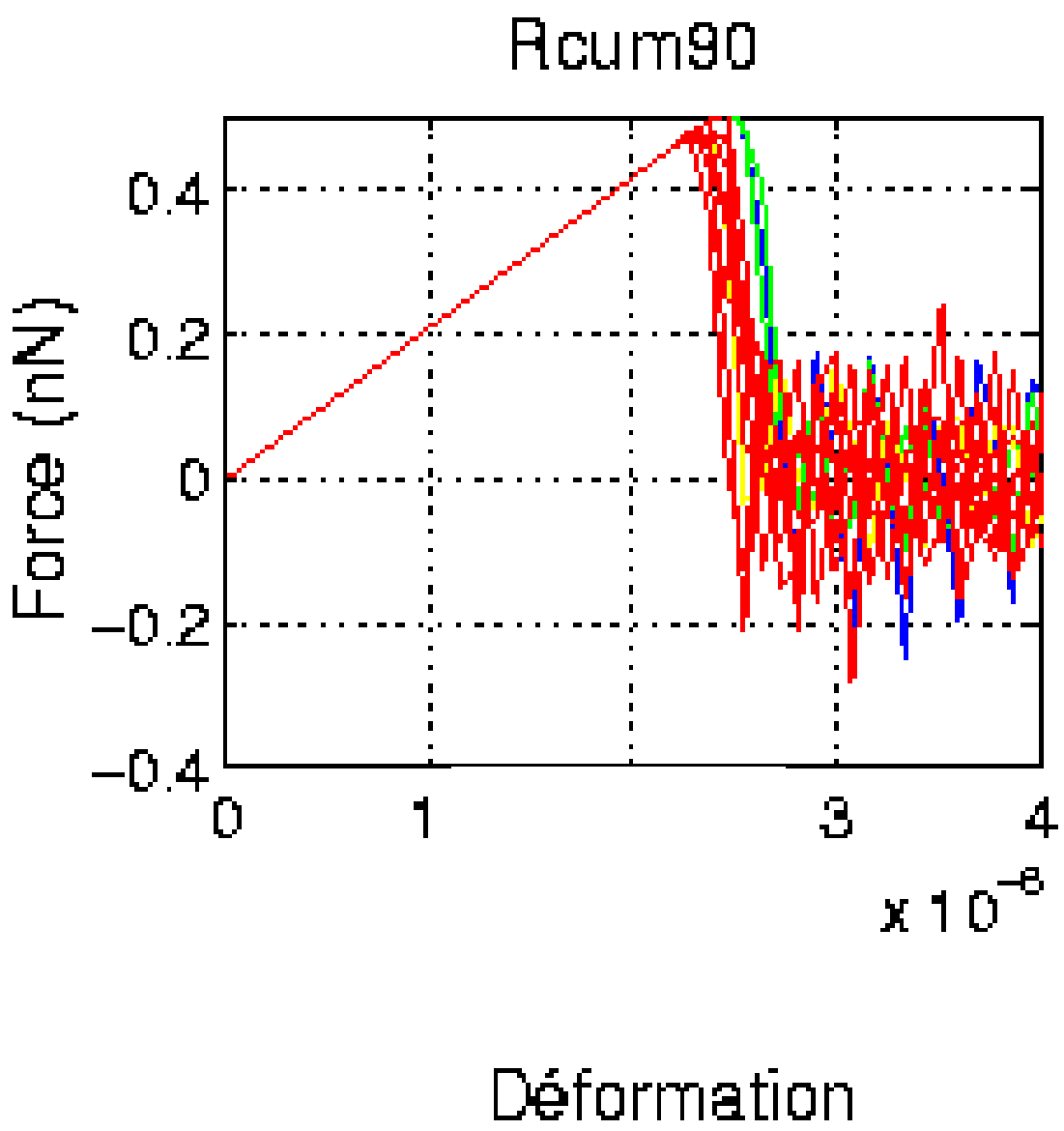

Figure 9

Strain

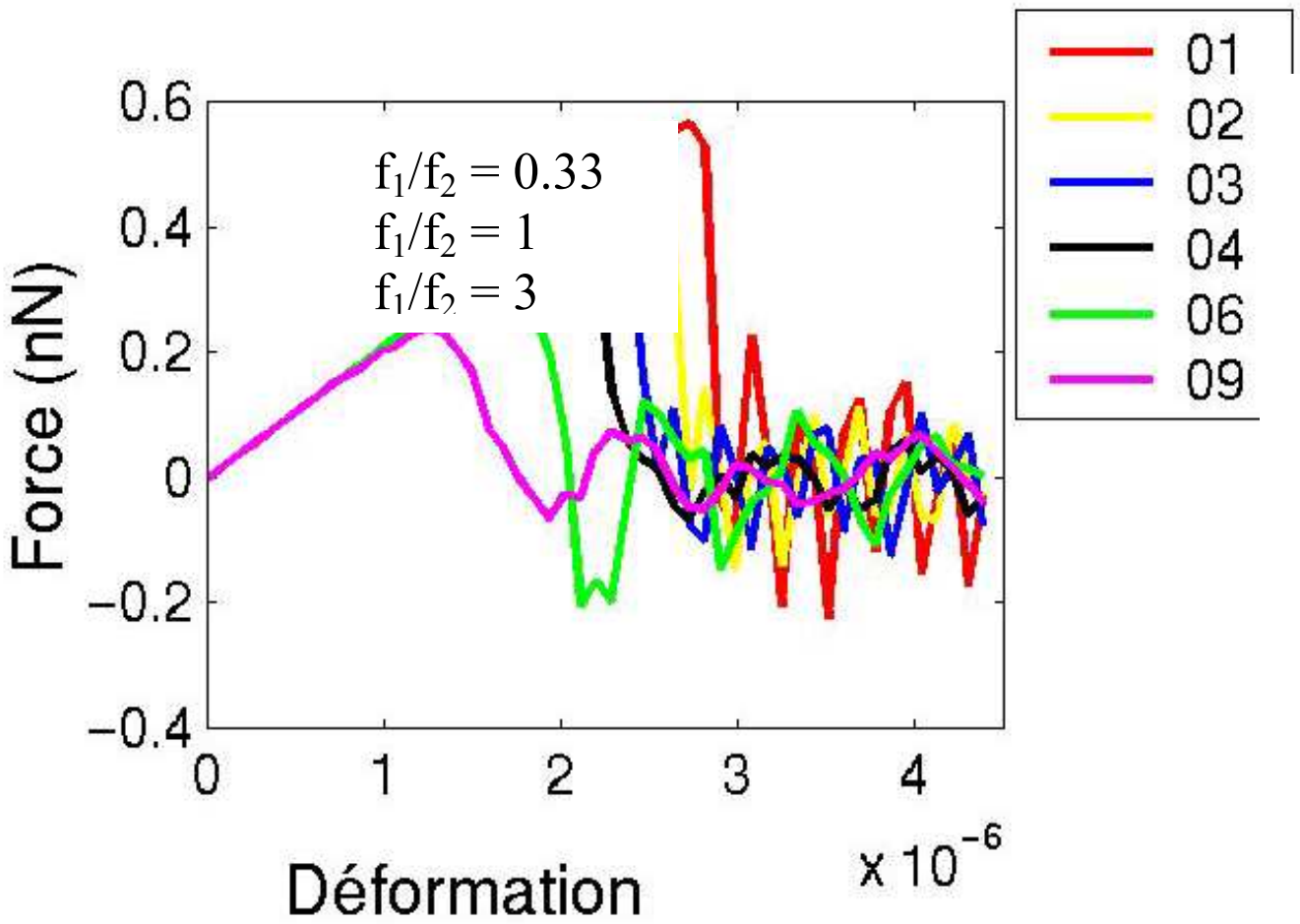

Figure 10

Strain

Strain

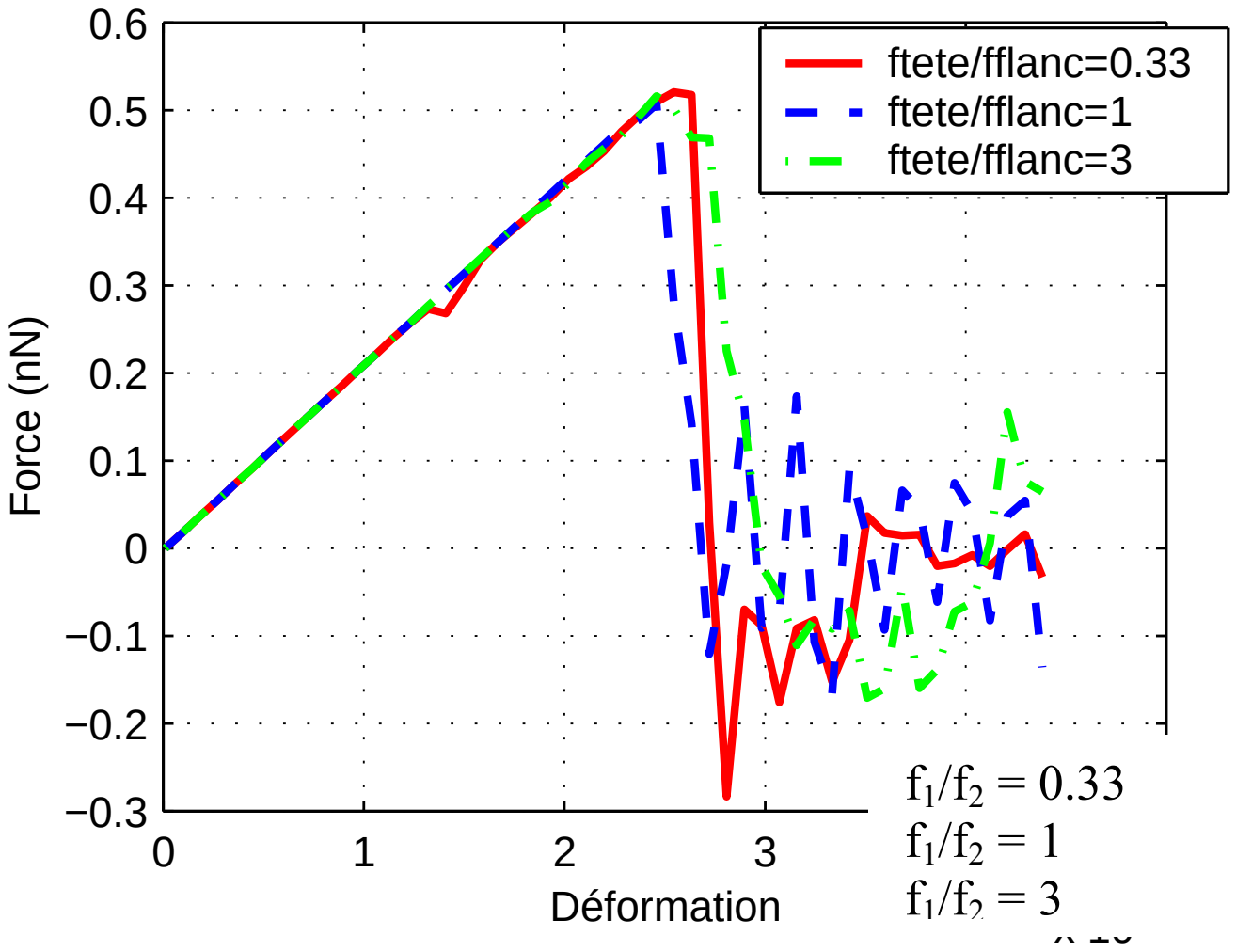

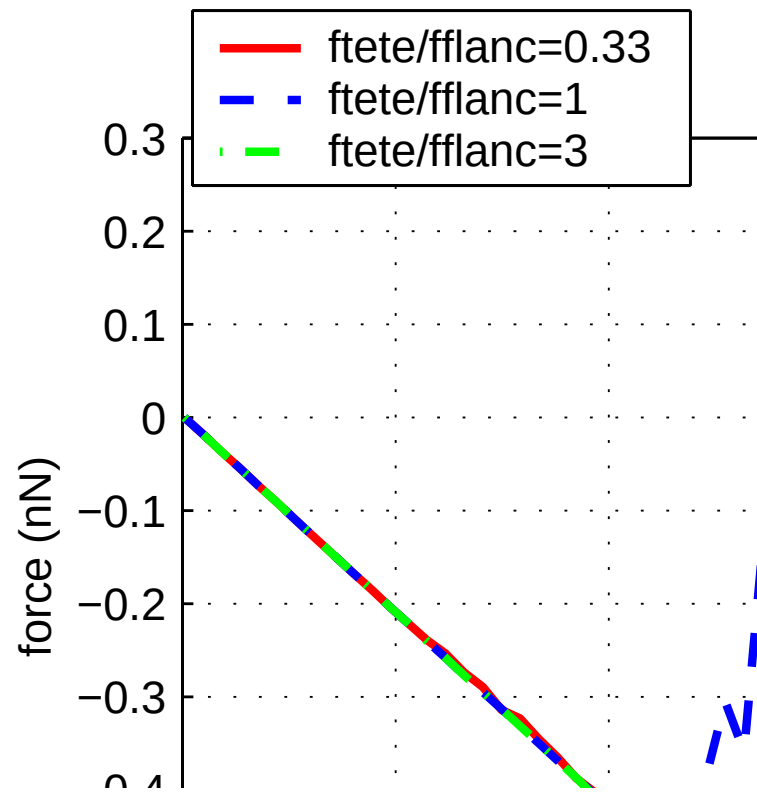

Figure 11

Strain

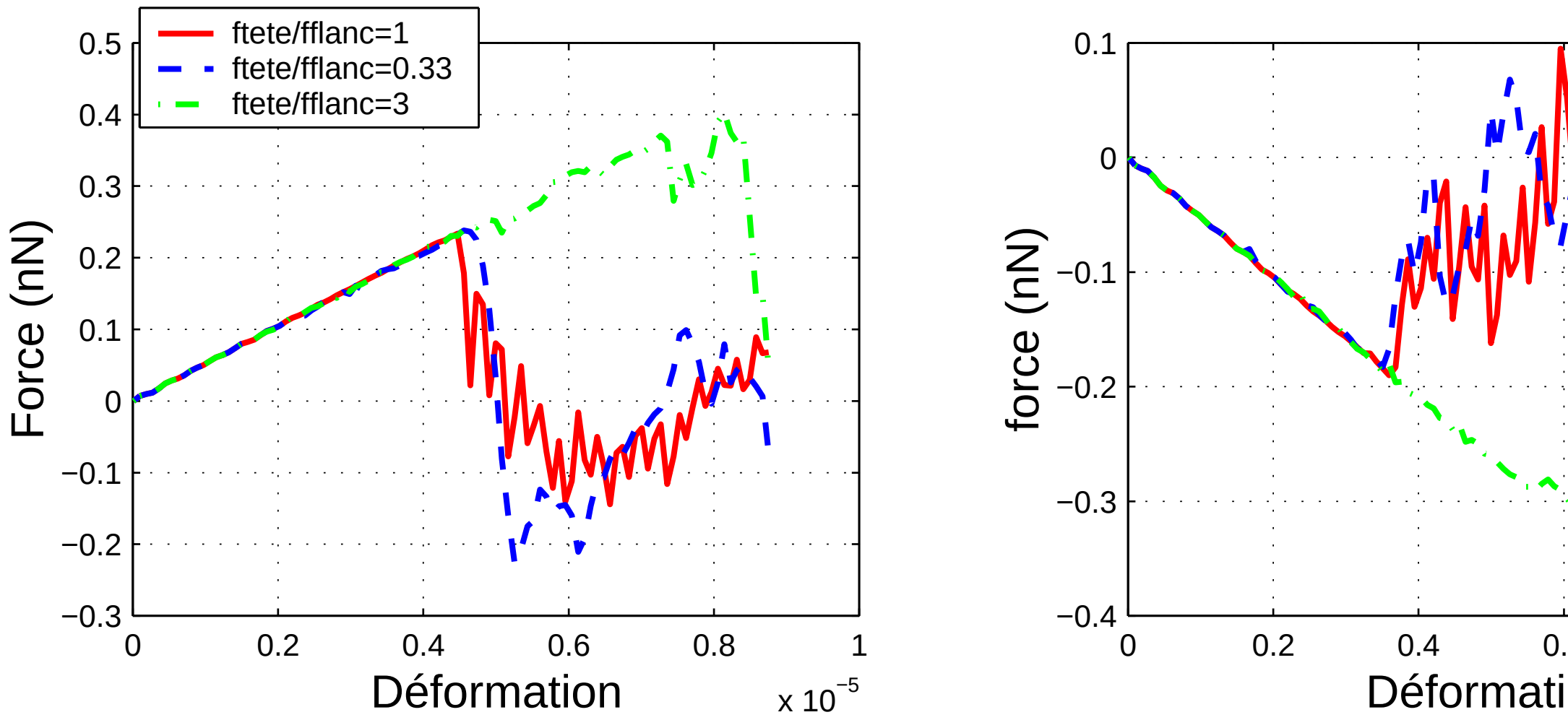

Figure 12

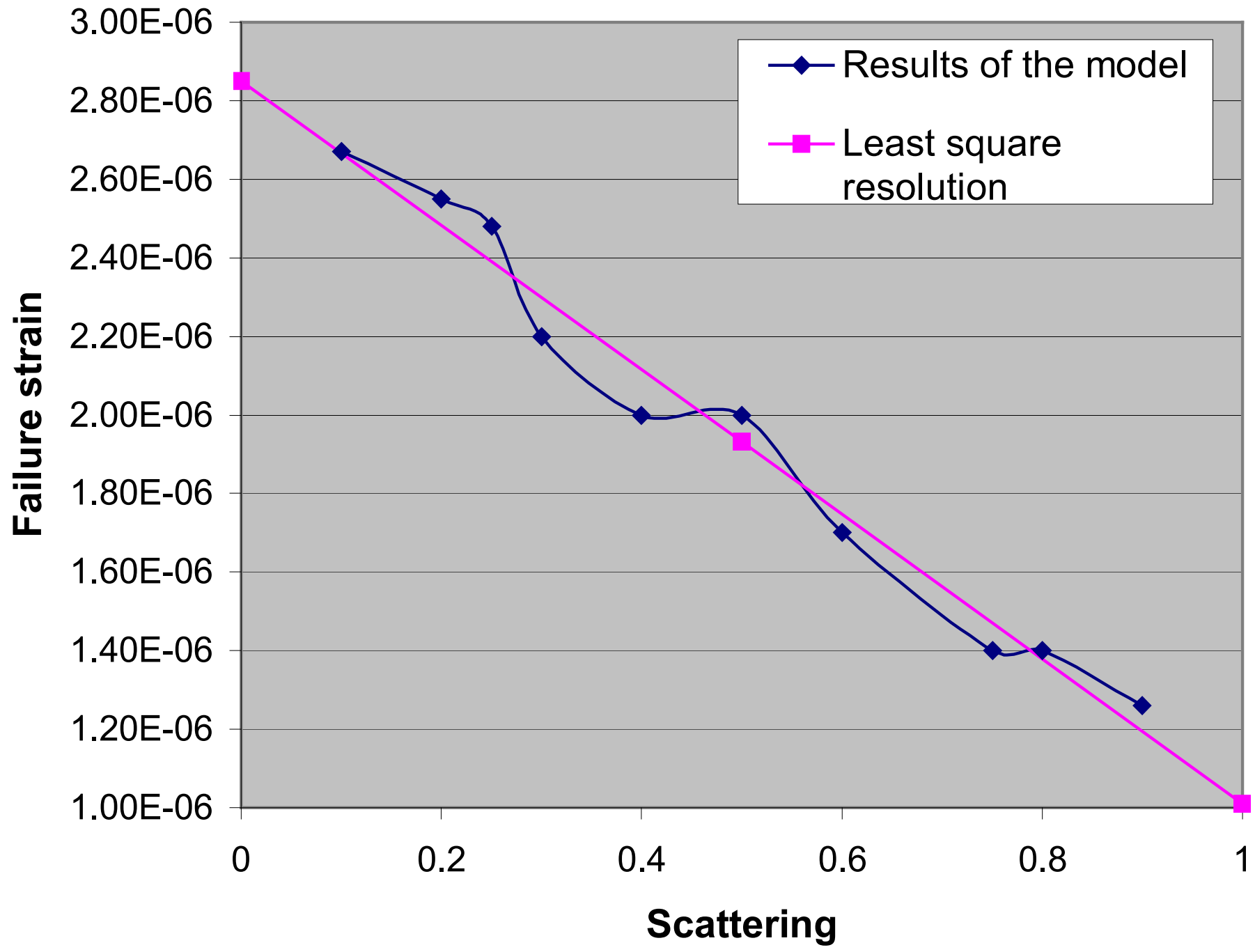

Figure 13

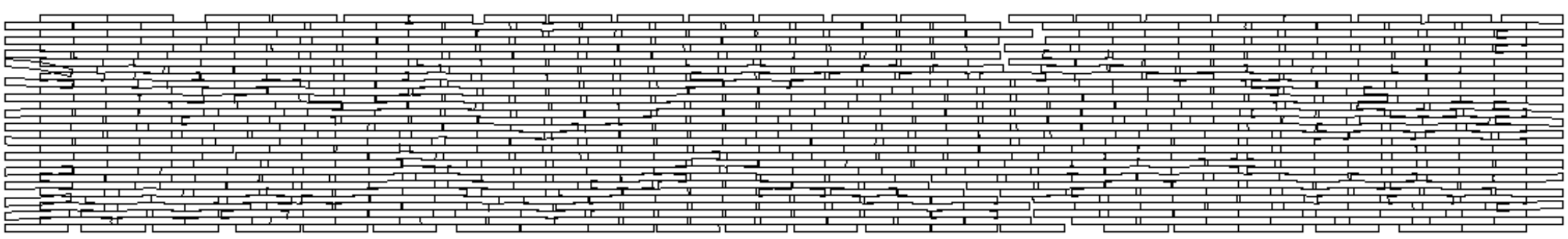

(a)

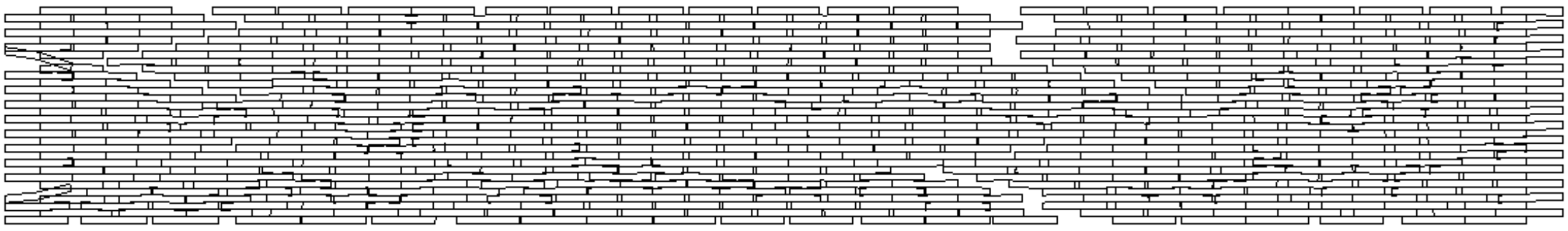

(b)

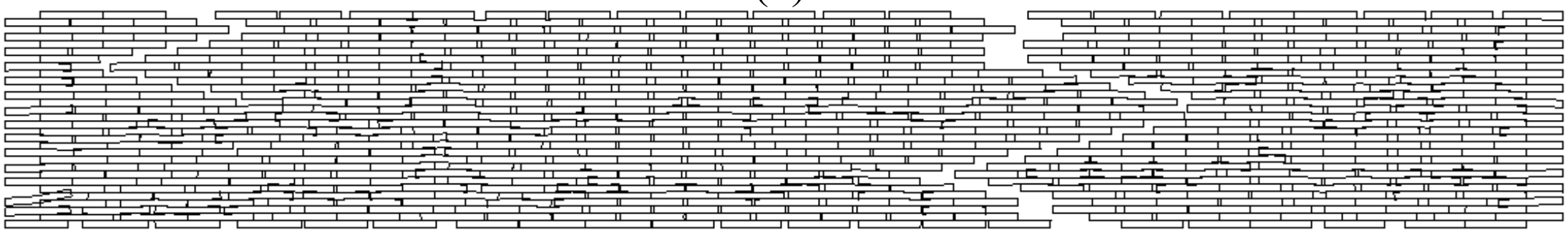

(c)

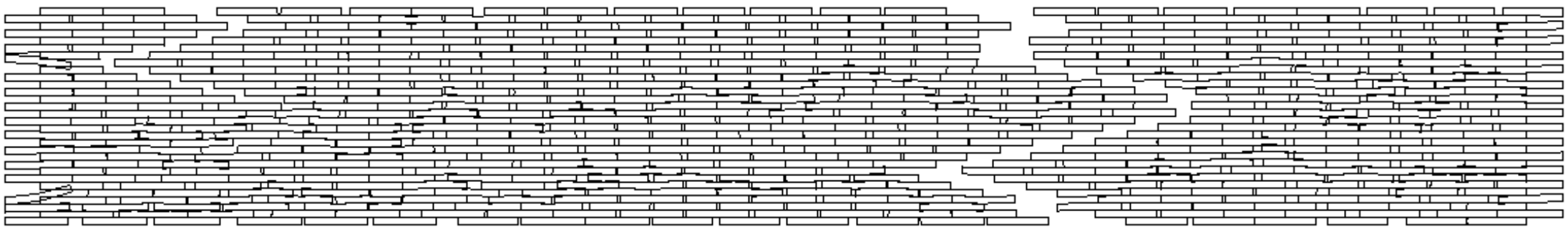

(d)

Figure 14

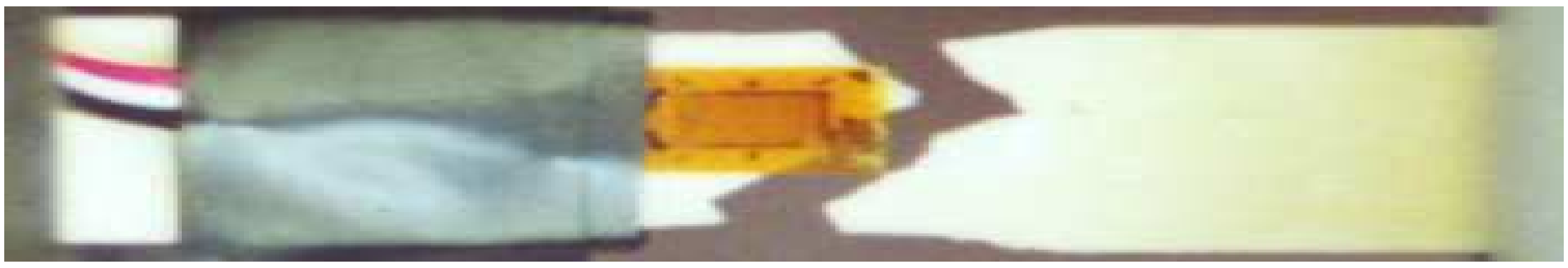

Figure 15

Table I

| | **Failure stress (nN/μm²)** |
|---|---|
| **21 fibres longitudinally in 30 successive layers** | 0.26 |
| **7 fibres longitudinally in 90 successive layers** | 0.25 |
| **42 fibres longitudinally in 60 successive layers** | 0.26 |